\documentclass[12pt]{article}
\usepackage{amsmath,amssymb,amsfonts}

\newtheorem{theorem}{Theorem}[section]
\newtheorem{lemma}[theorem]{Lemma}
\newtheorem{proposition}[theorem]{Proposition}

\newcommand{\cN}{\mathcal{N}}
\newcommand{\cB}{\mathcal{B}}
\newcommand{\ol}{\overline}
\newcommand{\wt}{\widetilde}

\begin{document}

\title{Nonlinear Averaging in Economics}
\author{V.~P.~Maslov\thanks{Moscow Institute of Electornics and Mathematics,
pm@miem.edu.ru}}
\date{}

\maketitle

\abstract{Kolmogorov nonlinear averaging is complemented by a
natural axiom. 
 For this averaging, we prove a theorem on large deviations as
well as establish the relationship to the tunnel canonical operator}

\section{Kolmogorov nonlinear averaging}
 A sequence of functions~$M_n$
defines a {\it regular\/} type
of average if the following conditions are satisfied (Kolmogorov):
\begin{description}
\item{I.}
$M(x_1,x_2,\dots,x_n)$
is continuous and monotone in each variable; to be definite,
we assume that~$M$
is increasing in each variable;
\item{II.}
$M(x_1,x_2,\dots,x_n)$
is
a symmetric function%
\footnote{In
our case, symmetry follows from the Bose statistics of banknotes.};
\item{III.}
the average of identical numbers is equal to their common value:
$M(x,x,\dots,x)=  x$;
\item{IV.}
a group of values can be replaced by their own average, without
changing the overall average:
$$
M(x_1,\dots,x_m,y_1,\dots,y_n)
=M_{n+m}(x,\dots,x,y_1,\dots,y_n),
$$
where
$x=M(x_1,\dots,x_m)$.
\end{description}

\begin{theorem}[Kolmogorov]
 If conditions {\rm I--IV} are satisfied, the average
$M(x_1,x_2,\dots, x_n)$
is of the form
\begin{equation}
M(x_1,x_2,\dots,x_n)
=\psi\biggl(\frac{\varphi(x_1)+\varphi(x_2)+\dots+
\varphi(x_n)}n\biggr),
\tag{1}
\end{equation}
where
$\varphi$
is a continuous, strictly monotone function and~$\psi$
is its inverse.
\end{theorem}

 For the proof of Theorem~1, see~\cite{1}.

\section{Main axiom of averaging}
 For a stable system, it is fairly obvious that the following axiom
must be satisfied:
\begin{description}
\item{V.}
if the same quantity~$\omega$
is added to each~$x_k$,
then
the average will increase by the same amount~$\omega$.
\end{description}
 I have had a detailed discussion of this axiom with the
practicing economist V.~N.~Baturin.

 Obviously, in normal conditions, the nonlinear average of~$x_i$
must also increase by this amount.
 We take this fact as {\it Axiom\/}~V.

 This axiom leads to a unique solution in the nonlinear case,
i.e., it is naturally satisfied in the linear case (the arithmetic
mean) and by a unique (up to an identical constant by which we
can multiply all the incomes~$x_i$)
nonlinear function.

 In fact, the incomes~$x_i$
are calculated in some currency
and, in general, must be multiplied by some constant~$\beta$
corresponding to the purchasing power of this currency, so that
this constant (the parameter~$\beta$)
must be incorporated
into the definition of an income.
 Therefore, we assert that there
exists a unique nonlinear function satisfying Axiom~V.

 The function
$\varphi(x)$
is of the form
\begin{equation}
\varphi(x)=C\exp(Dx)+B,
\tag{2}
\end{equation}
where
$C,D\ne  0$,
and
$B$
are numbers independent of
$x$.

\section{Semiring: an example of a linear self-adjoint operator}
 Consider the semiring generated by the nonlinear average and the
space~$L_2$ with values in this semiring.

 First, consider the heat equation
\begin{equation}
\frac{\partial u}{\partial t}
=\frac h2\,\frac{\partial^2u}{\partial x^2}.
\tag{3}
\end{equation}
 Here~$h$
is a small parameter, whose smallness is not
needed at this point.

 Equation~\thetag{3} is linear.
 As is well known, this means that
if $u_1$ and~$u_2$
are its solutions, then their linear combination
\begin{equation}
u=\lambda_1u_1+\lambda_2u_2
\tag{4}
\end{equation}
is also its solution.
 Here~$\lambda_1$
and~$\lambda_2$
are constants.

 Next, we make the following substitution.
 Let
\begin{equation}
u=e^{-M/h}.
\tag{5}
\end{equation}
 Then the new unknown function
$M(x,t)$
satisfies the nonlinear
equation
\begin{equation}
\frac{\partial M}{\partial t}
+\frac12\biggl(\frac{\partial M}{\partial x}\biggr)^2
-\frac h2\,\frac{\partial^2M}{\partial x^2}=0.
\tag{6}
\end{equation}
 This well known equation is sometimes called
the {\it B\"urgers equation\/}%
\footnote{The ordinary B\"urgers equation
is obtained from~(6) by differentiating with respect to~$x$
and substituting
$v=\partial M/\partial x$.}.

 To the solution~$u_1$
of Eq.~\thetag{3} there corresponds
the solution
$M_1=-h\ln u_1$
of Eq.~\thetag{6} and to the solution~$u_2$
of Eq.~\thetag{3} there corresponds the solution
$M_2=-h\ln u_2$
of Eq.~\thetag{6}.
 To the solution~\thetag{4} of Eq.~\thetag{3}
there corresponds the solution
$$
M=-h\ln\bigl(e^{-(M_1+\mu_1)/h}+ e^{-(M_2+\mu_2)/h}\bigr),
$$
where
$\mu_i=-h\ln\lambda_i$,
$i=1,2$.

 Hence Eq.~\thetag{6} is also linear, but it is linear in the function
space with the following operations:
\begin{description}
\item{$\bullet$}
the operation of addition
$a\oplus b=-h\ln(e^{-a/h}+  e^{-b/h})$;
\item{$\bullet$}
the operation of multiplication
$a\odot\lambda=a+  \lambda$.
\end{description}
 Now, the substitution
$M=-h\ln u$
takes zero to infinity and
one to zero.
 Thus, in this new space, the generalized zero is~$\infty$:
$\bold0=\infty$,
and the generalized unit is the ordinary zero:
$\bold1= 0$.
 The function space in which the operations~$\oplus$
and~$\odot$
have been introduced together with the zero
$\bold0$
and the unit
$\bold1$
is isomorphic to the ordinary function space
with ordinary multiplication and addition.

 We can thus imagine that, somewhere on another planet, people
have grown accustomed to the newly introduced operations~$\oplus$
and~$\odot$
and, from their point of view, Eq.~\thetag{6}
is linear.

 All this, of course, is trivial and there is no need for people
living on our planet to learn new arithmetical operations, because
we can, by a change of function, pass from Eq.~\thetag{6} to
Eq.~\thetag{3}, which is linear in the accepted meaning.
 However,
it turns out that the ``kingdom of crooked mirrors,'' which
this semiring yields, is related to ``capitalist'' economics~\cite{2}.

 In the function space with values in the ring
$a\oplus b=-h\ln(e^{-a/h}+  e^{-b/h})$,
$\lambda\odot b=\lambda+  b$,
we introduce the inner
product
$$
(M_1,M_2)=-h\ln\int e^{(M_1+W_2)/h}\,dx.
$$
 Let us show that it possesses bilinear properties in this space;
namely,
$$
(a\oplus b,c)=(a,c)\oplus(b,c),
\qquad
(\lambda\odot a,c)=\lambda\odot(a,c).
$$
 Indeed,
\begin{align}
(a\oplus b,c)
&=-h\ln\biggl(\int\exp\biggl(\frac{-(-h\ln(e^{-a/h}+e^{-b/h})+
c)}h\biggr)
dx\biggr)\tag{7}
\\ &
=-h\ln\biggl(\int(e^{-a/h}+e^{-b/h})e^{-c/h}\,dx\biggr)\notag
\\ &
=-h\ln\biggl(\int e^{-(a+c)/h}\,dx
+\int e^{-(b+c)/h}\,dx\biggr)
=(a,c)\oplus(b,c),\notag
\\
(\lambda\odot a,c)
&=-h\ln\int e^{-(a+\lambda)/h}e^{-c/h}\,dx\notag
\\ &
=-h\ln\biggl(e^{-\lambda/h}\int e^{-(a+c)/h}\,dx\biggr)
=\lambda+\ln\int e^{-(a+c)/h}\,dx
=\lambda\odot(a,c).\notag
\end{align}
 Let us give an example of a self-adjoint operator in this space.
 Consider the operator
$$
L\:W\to W\odot\biggl(-h\ln\biggl(\frac{(W')^2}{h^2}
-\frac{W''}h\biggr)\biggr).
$$
 And now let us verify its self-adjointness:
\begin{align}
&
(W_1,LW_2)
=-h\ln\int e^{-(W_1+LW_2)/h}\,dx
\tag{8}
\\ &\qquad
=-h\ln\int\exp\biggl[-\biggl(W_1+W_2-h\ln\biggl(\frac{W'_2)^2}{h^2}
-\frac{W''_2}h\biggr)\biggr)\bigg/h\biggr]dx\notag
\\ &\qquad
=-h\ln\int e^{-W_1/h}e^{-W_2/h}\biggl(\frac{(W_2')^2}{h^2}
-\frac{W''_2}h\biggr)dx
=-h\ln\int e^{-W_1/h}\frac{d^2}{dx^2}e^{-W_2/h}\,dx\notag
\\ &\qquad
=-h\ln\int\frac{d^2}{dx^2}
e^{-W_1/h}e^{-W_2/h}\,dx
=-h\ln\int e^{-W_1/h}\biggl(\frac{(W'_1)^2}{h^2}
-\frac{W''_1}h\biggr)e^{-W_2/h}\,dx\notag
\\ &\qquad
=-h\ln\int\exp\biggl[-\biggl(W_1-h\ln\biggl(\frac{(W'_1)^2}{h^2}
-\frac{W'_2}h\biggr)\biggr)\bigg/h\biggr]dx\notag
\\ &\qquad
=-h\ln\int e^{-(LW_1+W_2)/h}\,dx
=(LW_1,W_2).\notag
\end{align}
 It is also easy to verify linearity.

 We construct the resolving operator of the B\"urgers equation:
$L\:W_0\to  W$,
where
$W$
is the solution of Eq.~\thetag{6}
satisfying the initial condition
$W|_{t=0}=  0$.

 The solution of Eq.~\thetag{3} satisfying the condition
$u|_{t=0}=  u_0$
is of the form
$$
u=\frac1{\sqrt{2\pi h}}
\int e^{-(x-\xi)^2/2th}u_0(\xi)\,d\xi.
$$
 Taking into account the fact that
$u=  e^{-W/h}$,
$W=-h\ln u$,
we obtain the resolving operator~$L_t$
of the B\"urgers equation:
\begin{equation}
L_tW_0=-\frac h{\sqrt{2\pi h}}
\ln\int e^{-((x-\xi)^2/2th+M(\xi)/h)}\,d\xi.
\tag{9}
\end{equation}
 The operator~$L_t$
is self-adjoint in terms of the new inner
product.

\section{Theorem on the nonlinear average}
 Consider a collection of prices~$\lambda_i$,
where
$i=1,\dots,n$,
and a collection of numbers~$g_i$
equal to the number of financial
instruments (FI), which are shares, bonds, etc., of {\it different
types\/} having the price~$\lambda_i$.
 The prices~$\lambda_i$
are, by definition, positive numbers; without loss of generality,
we can further assume that the prices are numbered so as to
satisfy the inequalities
\begin{equation}
0<\lambda_1<\lambda_i<\lambda_n
\qquad \text{for
all}\quad i=2,\dots,n-1.
\tag{10}
\end{equation}
 The total number of FIs of different type is denoted by~$G$.
 This number is
\begin{equation}
G=\sum_{i=1}^ng_i.
\tag{11}
\end{equation}
 Let~$k_i$
denote the number of FIs purchased at the price~$\lambda_i$.
 Since
$g_i$
different FIs are sold at the price~$\lambda_i$,
it follows that the number of different methods of buying~$k_i$
 FI's at the price~$\lambda_i$
can be expressed by
\begin{equation}
\gamma_i(k_i)=\frac{(k_i+g_i-1)!}{k_i!\,(g_i-1)!}.
\tag{12}
\end{equation}
 The number of different methods of buying the collection~$\{k\}$
of FIs consisting of
$k_1,\dots,k_n$
 FIs bought at corresponding
prices
$\lambda_1,\dots,\lambda_n$,
is expressed by~\thetag{12}
and is equal to
\begin{equation}
\gamma(\{k\})=\prod_{i=1}^n\gamma_i(k_i)
=\prod_{i=1}^n\frac{(k_i+g_i-1)!}{k_i!\,(g_i-1)!}.
\tag{13}
\end{equation}
 The expenditure of the buyer in the purchase of the collection
FI
$\{k\}$
is equal to
\begin{equation}
\cB(\{k\})=\sum_{i=1}^n\lambda_ik_i.
\tag{14}
\end{equation}

 Let us carry out nonlinear averaging of the expenditure~\thetag{14}
over budget restraints:
\begin{equation}
M(\beta,N)
=-\frac1{\beta N}\ln\biggl(\frac{N!\,(G-1)!}{(N+G-1)!}
\sum_{\{k_i=N\}}\gamma(\{k\})\exp\bigl(-\beta\cB(\{k\})\bigr)\biggr),
\tag{15}
\end{equation}
where
$\beta$
is a positive parameter.

 We split the FIs arbitrarily into~$m$
nonintersecting groups,
where
$m\le  n$.
 This implies that, by employing some
method, we choose two sequences~$i_\alpha$
and~$j_\alpha$,
where
$\alpha=1,\dots,m$,
satisfying the conditions
\begin{equation}
i_\alpha\le j_\alpha,\; \;
i_{\alpha+1}=j_\alpha+1, \quad \alpha=1,\dots,m,
\qquad
i_1=1,\; \;j_m=n,
\tag{16}
\end{equation}
and assume that the FIs are contained in the group indexed by~$\alpha$
if the number~$i$
of its price~$\lambda_i$
satisfies
the condition
$i_\alpha\le i\le  j_\alpha$.
 Note that
there are a number of ways by which we can choose the
sequences~$i_\alpha$
and~$j_\alpha$
satisfying conditions~\thetag{16}.

 The number of FIs contained in the group indexed by~$\alpha$
is expressed by the formula
\begin{equation}
G_\alpha=\sum_{i=i_\alpha}^{j_\alpha}g_i,
\tag{17}
\end{equation}
while the number of FIs bought from the group with number~$\alpha$
in the purchase of the collection
$\{k\}=k_1,\dots,k_n$,
is expressed
by the formula
\begin{equation}
N_\alpha=\sum_{i=i_\alpha}^{j_\alpha}k_i.
\tag{18}
\end{equation}
 The collection
$\{k\}$
satisfies condition~\thetag{31}; therefore,
the collection of numbers~$N_\alpha$
satisfies the equality
\begin{equation}
\sum_{\alpha=1}^mN_\alpha=N;
\tag{19}
\end{equation}
also, it follows from~\thetag{11} that the number~$G_\alpha$
satisfies the relation
\begin{equation}
\sum_{\alpha=1}^mG_\alpha=G.
\tag{20}
\end{equation}
 Here~$G$
depends on~$N$
so that the following relation
holds:
\begin{equation}
\lim_{N\to\infty}\frac GN=\wt g>0.
\tag{21}
\end{equation}
 In addition, we assume that the partition of the FIs into
the groups~\thetag{16}
satisfies the following condition:
$m$
is independent of
$N$,
and the~$G_\alpha$
depend on~$N$
so that the following
relations hold:
\begin{equation}
\lim_{N\to\infty}\frac{G_\alpha}N
=\wt g_\alpha>0
\qquad \text{for
all}\quad
\alpha=1,\dots,m.
\tag{22}
\end{equation}
 Hence
$$
\lim_{N\to\infty}\frac{N_\alpha}N\approx\ol n_\alpha>0,
\qquad \sum\ol n_\alpha=1.
$$
 By~\thetag{20}, the quantities~$\wt g_\alpha$
and~$\wt g$
are related by
\begin{equation}
\sum_{\alpha=1}^m\wt g_\alpha=\wt g.
\tag{23}
\end{equation}
 We denote
\begin{equation}
\cN_\alpha(\beta,N)
=\sum_{i=i_\alpha}^{j_\alpha}
\frac{g_i}{\exp(\beta(\lambda_i+\nu))-1},
\tag{24}
\end{equation}
where
$\delta>  0$
is an arbitrary parameter and~$\nu$
is specified by the equation
\begin{equation}
N=\sum_{i=1}^n\frac{g_i}{\exp(\beta(\lambda_i+\nu))-1}.
\tag{25}
\end{equation}
 Further, we use the notation
\begin{equation}
\Gamma(\beta,N)
=\frac{(N+G-1)!}{N!\,(G-1)!}
\exp\bigl(-\beta NM(\beta,N)\bigr).
\tag{26}
\end{equation}

 We can also consider the case
$\beta<  0$;
moreover, for
$\beta<  0$,
we choose the solution of Eq.~\thetag{25} satisfying the condition
$\nu<  -\lambda_n$
and, for
$\beta>  0$,
we choose the solution satisfying the condition
$\nu>  -\lambda_1$.

\begin{theorem}
 Let condition~\thetag{22} be satisfied, and let
$\Delta=aN^{3/4+\delta}$,
where
$a$
and
$\delta<1/3$
are positive parameters independent of
$N$.
 Then, for any
$\varepsilon>  0$,
the following relation holds as
$N\to  \infty$,
regardless of the value of~$N${\rm:}
\begin{align}
&
\frac1{\Gamma(\beta,N)}
\sum_{\sum\{k_i\}=N,\,\sum^m_{\alpha=1}
(N_\alpha(\{k\})-\cN_\alpha(\beta,N))^2\ge\Delta}
\gamma(\{k\})\exp\bigl(-\beta\cB(\{k\})\bigr)
\tag{27}
\\ &\qquad
=O\biggl(\exp\biggl(-\frac{(1-\varepsilon)a^2N^{1/2+2\delta}}
{2\wt gd}\biggr)\biggr),
\notag
\end{align}
where
$d$
is defined by
\begin{equation}
d=\begin{cases}
\dfrac{\exp(-\beta(\lambda_1+\nu))}{(\exp(-\beta(\lambda_1+\nu))-1)^2}
& \text{for}\ \beta<0, \\
\dfrac{\exp(-\beta(\lambda_n+\nu))}{(\exp(-\beta(\lambda_n+\nu))-1)^2}
& \text{for}\ \beta>0.
\end{cases}
\tag{28}
\end{equation}
 In other words, the contribution to the average expenditure of
the buyer from the number~$N_\alpha$
of the purchased FI's
{\rm(}of the order of~$N${\rm)}, which
differs from
$\cN_\alpha(\beta,N)$
by a value
$O(N^{3/4+\delta})$,
is of exponentially small value.
\end{theorem}

 Thus, for the case in which the FIs can be divided into sufficiently
large groups (condition~\thetag{22}), given a fixed nonlinear
average expenditure of the buyer, the number of purchased FIs
belonging
to the group~$\alpha$
is, in most cases, approximately equal to
$\cN_\alpha(\beta,N)$,
which is an analog of the law of large numbers.

 To find~$\beta$,
we propose the following method.
 Suppose
we are given a collection~$N_\alpha^a$
of actual purchases from the group
$\alpha$,
$\alpha=1,2,\dotsb,m$,
of financial instruments.
 Consider the following function of~$\beta$:
\begin{equation}
DN(\beta)=\sum_{\alpha=1}^m
\bigl(N_\alpha^a-\cN_\alpha(\beta,\nu_\alpha)\bigr)^2,
\tag{29}
\end{equation}
where
$\cN_\alpha(\beta,\nu_\alpha)$
is given by~\thetag{24}.
 For~$\beta$
we take the value for which the function~\thetag{29}
attains its minimum.

 The minimum of this value, which can be specified, for example,
at the stock exchange, shows the degree of randomness of the
distribution
of FIs over the portfolios of the traders.
 If the minimum is zero,
then the distribution over the portfolios is random (see~\cite{3}).

 Before proving the theorem, we shall prove several assertions
and lemmas.

 Suppose that the buyer possesses a sum of money~$B$
(budget restraint), but not all of this sum is necessarily
spent on purchases; suppose that the buyer purchases~$N$~FIs.
 The total number of different types of purchase corresponding to the~$N$
purchased~FIs
when the sum of money spent does not exceed~$B$
can be expressed by the formula~\cite{4}
\begin{equation}
\sigma(B,N)
=\sum_{\{k=N\}}\gamma(\{k\})\Theta\bigl(B-\cB(\{k\})\bigr).
\tag{30}
\end{equation}
 Here
$\cB(\{k\})\le  B$,
\
$\cB(\{k\})$
is defined by~\thetag{14},
$\Theta(x)$
is the Heaviside function:
$$
\Theta(x)=\begin{cases}
0& \text{for}\ x<0,\\
1& \text{for}\ x\ge1,
\end{cases}
$$
and~$\sum_{\{k\}}^N$
denotes the sum over all collections
$\{k\}$
of nonnegative integers
$k_1,\dots,k_n$
satisfying the
condition
\begin{equation}
\sum_{i=1}^nk_i=N.
\tag{31}
\end{equation}
 Note that, by~\thetag{10}, the expenditure~\thetag{14} for any
collection
$\{k\}$
fulfilling condition~\thetag{31} satisfies
the inequalities
\begin{equation}
0<N\lambda_1\le \cB(\{k\})\le N\lambda_n.
\tag{32}
\end{equation}

 The number
$\sigma(B,N)$
~\thetag{30} of different types of purchase
is a piecewise constant function of the variable~$B$
and possesses the following properties:
$$
\alignedat2
\sigma(B,N)
&\le\sigma(B',N) &\qquad \text{for}\quad B&\le B',
\\
\sigma(B,N)
&=0 &\qquad \text{for}\quad B&<N\lambda_1,
\\
\sigma(B,N)
&=\sigma(N\lambda_n,N) &\qquad \text{for}\quad B&\ge N\lambda_n.
\endalignedat
$$
 These properties of
$\sigma(B,N)$
follow immediately from the definition~\thetag{30}
and inequality~\thetag{32}.

\begin{proposition}
 For any positive~$\beta$
and any~$N$,
the following identity
holds{\rm:}
\begin{equation}
\int_0^\infty e^{-\beta B}\,d\sigma(B,N)
=\Gamma(\beta,N).
\tag{33}
\end{equation}
\end{proposition}

{\bf Proof.}
 Substituting~\thetag{30} into the left-hand side of
relation~\thetag{33}
and changing the order of summation and integration, we obtain
\begin{equation}
\int_0^\infty e^{-\beta B}\,d\sigma(B,N)
=\sum_{\{k\}}^N\gamma(\{k\})\int_0^\infty
e^{-\beta B}\,d\Theta\bigl(B-\cB(\{k\})\bigr).
\tag{34}
\end{equation}
 Next, we use the equality
\begin{equation}
\int_0^\infty e^{-\beta B}\,d\Theta(B-B_0)
=e^{-\beta B_0},
\tag{35}
\end{equation}
which is valid for all positive~$\beta$
and~$B_0$.
 Taking~\thetag{35}
into account, from~\thetag{34} we obtain the following equality:
\begin{equation}
\int_0^\infty e^{-\beta B}\,d\sigma(B,N)
=\sum_{\{k\}}^N\gamma(\{k\})
\exp\bigl(-\beta\cB(\{k\})\bigr).
\tag{36}
\end{equation}
 It follows from~\thetag{15} and~\thetag{26} that the expression
on the right-hand side of relation~\thetag{36} is equal to
$\Gamma(\beta,N)$.
 Thus, the proposition is proved.

 We introduce the following function
$\zeta(\beta,\nu)$
defined
for all
$\nu>  -\lambda_1$
and
$\beta>  0$:
\begin{equation}
\zeta(\beta,\nu)
=\sum_{N=0}^\infty\Gamma(\beta,N)e^{-\beta N\nu}.
\tag{37}
\end{equation}

\begin{proposition}
 For all
$\nu>  -\lambda_1$
and
$\beta>  0$,
the following relation holds{\rm:}
\begin{equation}
\zeta(\beta,\nu)
=\prod_{i=1}^n\frac1{(1-\exp(-\beta(\lambda_i+\nu)))^{g_i}}.
\tag{38}
\end{equation}
\end{proposition}

{\bf Proof.}
 By formulas~\thetag{15}, \thetag{26}, we have
\begin{equation}
\sum_{N=0}^\infty\Gamma(\beta,N)e^{-\beta N\nu}
=\sum_{N=0}^\infty e^{-\beta N\nu}
\sum_{\{k\}}^N\gamma(\{k\})\exp\bigl(-\beta\cB(\{k\})\bigr).
\tag{39}
\end{equation}
 Taking into account the explicit form of~$\gamma(\{k\})$~\thetag{13}
and~$\cB(\{k\})$ (see~\thetag{14}), we find that the sum in
relation~\thetag{39}
splits into the following product of sums:
\begin{equation}
\sum_{N=0}^\infty e^{-\beta N\nu}
\sum_{\{k\}}^N\gamma(\{k\})\exp\bigl(-\beta\cB(\{k\})\bigr)
=\prod_{i=1}\biggl(\sum_{k_i=1}^\infty
\frac{(k_i+g_i-1)!}{k_i!\,(g_i-1)!}
e^{-\beta(\lambda_i+\nu)k_i}\biggr).
\tag{40}
\end{equation}
 We now use the equality
\begin{equation}
\sum_{k=0}^\infty\frac{(k+g-1)!}{k!\,(g-1)!}x^k
=\frac1{(1-x)^g},
\tag{41}
\end{equation}
valid for any natural number~$g$
and any~$x$
such that
$|x|<  1$.
 Since
$\nu>  -\lambda_1$
and
$\beta>  0$,
by~\thetag{41} from~\thetag{37}, \thetag{39}, \thetag{40}
we obtain~\thetag{38}.
 The assertion is proved.

 In order to prove certain properties of the function
$\Gamma(\beta,N)$ (see~\thetag{26}),
we shall use the following lemma.

\begin{lemma}
 The functions
$z_n(x_1,\dots,x_n;N)$
given by
\begin{equation}
z_l(x_1,\dots,x_l;N)
=\sum_{\{k\}}^N\exp\biggl(\sum_{i=1}^lx_ik_i\biggr),
\tag{42}
\end{equation}
where~$l$
and~$N$
are arbitrary integers satisfying the
conditions
$l\ge  2$
and
$N\ge  1$,
satisfy,
for all real numbers
$x_1,\dots,x_n$,
the inequality
\begin{equation}
z_l(x_1,\dots,x_l,N)^2
>z_l(x_1,\dots,x_l,N-1)z_l(x_1,\dots,x_l,N+1).
\tag{43}
\end{equation}
\end{lemma}

{\bf Proof.}
 Let us prove the lemma by induction.
 First,
we prove that~\thetag{43} is valid for
$l=  2$.
 If
$l=  2$
and
$x_2=  x_1$,
then, for all nonnegative integers~$N$,
the function~\thetag{42} is equal to
$$
z_2(x_1,x_1;N)
=\sum_{k=0}^Ne^{x_1N}=(N+1)e^{x_1N},
$$
and it can be verified by elementary means that, in that case,
inequality~\thetag{43} holds.
 If
$l=  2$
and
$x_2\ne  x_1$,
then, for all nonnegative integers~$N$,
the function~\thetag{42}
is of the form
\begin{equation}
z_2(x_1,x_2;N)
=\sum_{k=0}^Ne^{x_1k+x_2(N-k)}
=e^{x_2N}\frac{1-e^{(x_1-x_2)(N+1)}}{1-e^{x_1-x_2}}.
\tag{44}
\end{equation}
 Substituting~\thetag{44} into~\thetag{43}, after a few manipulations
we find that, in the case under consideration, inequality~\thetag{43}
is equivalent to the following inequality:
\begin{equation}
\frac12\bigl(e^{x_1-x_2}+e^{x_2-x_1}\bigr)>1.
\tag{45}
\end{equation}
 However, since the function
$y(x)=\exp(x)$
is convex, inequality~\thetag{45}
holds for all numbers~$x_1$
and~$x_2$
not equal to one
another.
 Thus, inequality~\thetag{43} is proved for
$l=  2$.
 Next, suppose that inequality~\thetag{43} is proved for
$l=  k$,
where
$k\ge  2$,
and let us prove that this inequality
holds for
$n=l+  1$.
 Note that the functions~\thetag{42}
possess the following property, which is a trivial consequence
of their explicit form:
\begin{equation}
z_{k+1}(x_1,\dots,x_{k+1};N)
=e^{x_{k+1}N}\sum_{m=0}^Nz_k(\wt x_1,\dots,\wt x_k;m),
\tag{46}
\end{equation}
where
$\wt x_i=x_i-  x_{k+1}$,
\
$i=1,\dots,n$.
 Substituting~\thetag{46} into inequality~\thetag{43} and carrying
out a few manipulations, we see that inequality~\thetag{43} for
$n=k+  1$
is equivalent to the inequality
\begin{equation}
z_k(\wt x;0)z_k(\wt x;N)
+\sum_{m=0}^{N-1}\bigl(z_k(\wt x;k+1)z_k(\wt x;N)
-z_k(\wt x;k)z(\wt x;N+1)\bigr)
>0,
\tag{47}
\end{equation}
where
$z_k(\wt x;m)$
denotes
$z_k(\wt x_1,\dots,\wt x_k;m)$.
 By assumption, inequality~\thetag{43} holds for
$n=  k$;
this implies that, for all
$k=0,\dots,N-  1$,
we have
\begin{equation}
z_k(\wt x;k+1)z_k(\wt x;N)-z_k(\wt x;k)z(\wt x;N+1)>0,
\tag{48}
\end{equation}
and hence all summands on the left-hand side of inequality~\thetag{47}
are positive, i.e., inequality~\thetag{47} holds.
 Thus, the validity
of~\thetag{43} for
$l=k+  1$
is proved and, therefore,
the assertion of the lemma is also proved by induction.

 Lemma~1 implies the convexity of the logarithm of the function
$\Gamma(\beta,N)$ (given by~\thetag{26}) with respect to the
discrete variable~$N$.

\begin{proposition}
 For all natural numbers~$N$
and all positive numbers~$\beta$,
the following inequalities hold{\rm:}
\begin{gather}
\Gamma(\beta,N)^2
>\Gamma(\beta,N-1)\Gamma(\beta,N+1),
\tag{49}
\\
\Gamma(\beta,N-1)
<e^{\beta\lambda_1}\Gamma(\beta,N).
\tag{50}
\end{gather}
\end{proposition}

{\bf Proof.}
 To prove these inequalities, it suffices to note that, in view
of formula~\thetag{15},
$\Gamma(\beta,N)$~\thetag{26} can be written
in the form
\begin{equation}
\Gamma(\beta,N)
=\sum_{\{k\}}^N\gamma(\{k\})\exp\bigl(-\beta\cB(\{k\})\bigr).
\tag{51}
\end{equation}
 The right-hand side of relation~\thetag{51} is equal to the
function~\thetag{42}
for
\begin{gather}
l=G,
\qquad
x_1=\dots=x_{g_1}=-\beta\lambda_1,
\quad
x_{g_1+1}=\dots=x_{g_1+g_2}=-\beta\lambda_2,
\quad \dots\,,\tag{52}
\\
x_{G-g_n+1}=\dots=x_G=-\beta\lambda_G;\notag
\end{gather}
this equality follows from the formula
$$
\sum_{0\le m_1,\dots,m_g}^{m_1+\dots+m_g=k}1
=\frac{(k+g-1)!}{k!\,(g-1)!}.
$$
 Thus, inequality
\thetag{49}
is a consequence of~\thetag{43} and, to prove~\thetag{50}, it
suffices to use relation~\thetag{46}.

 Property~\thetag{49} implies another property of
$\Gamma(\beta,N)$.

\begin{proposition}
 For any given positive~$\beta$
and for any nonnegative integer~$N$,
there exists a number
$\nu>  -\lambda_1$
such that, for all
$N'\ne  N$,
\begin{equation}
\Gamma(\beta,N)e^{-\beta N\nu}
>\Gamma(\beta,N')e^{-\beta N'\nu}.
\tag{53}
\end{equation}
\end{proposition}

{\bf Proof.}
 Using~\thetag{49}, we choose any~$\nu$
satisfying the inequality
\begin{equation}
\frac1\beta\ln\biggl(\frac{\Gamma(\beta,N+1)}{\Gamma(\beta,N)}\biggr)
<\nu
<\frac1\beta\ln\biggl(\frac{\Gamma(\beta,N)}{\Gamma(\beta,N-1)}\biggr).
\tag{54}
\end{equation}
 It follows from~\thetag{50} that such a~$\nu$ also
satisfies the
inequality
$\nu>  -\lambda_1$.
 Moreover, inequalities~\thetag{49}
and~\thetag{54} imply the inequalities
\begin{align}
e^{-\beta(N'+1)\nu}\Gamma(\beta,N'+1)
&>e^{-\beta N'\nu}\Gamma(\beta,N')
&\qquad \text{for
all}\quad N', \quad &0\le N'\le N-1,
\tag{55}
\\
e^{-\beta(N'+1)\nu}\Gamma(\beta,N'+1)
&<e^{-\beta N'\nu}\Gamma(\beta,N')
&\qquad \text{for
all}\quad N', \quad &N\le N'.\notag
\end{align}
 However, the validity of inequalities~\thetag{55} implies
that of inequality~\thetag{53} for all
$N'\ne  N$.
 The proposition is proved.

 Next, let us prove another lemma, which gives an estimate of the
nonlinear average~\thetag{15} via relation~\thetag{37}.
 Suppose that~$p_l$,
$l=0,1,\dots$,
is a sequence of numbers satisfying the conditions
\begin{equation}
p_l\ge0 \quad \text{for
all}\ l=0,1,\dots,
\qquad
\sum_{l=0}^\infty p_l=1,
\tag{56}
\end{equation}
and also
\begin{align}
\ol l
&\equiv\sum_{l=0}^\infty lp_l<\infty,
\tag{57}
\\
Dl
&\equiv\sum_{l=0}^\infty(l-\ol l)^2p_l<\infty.
\tag{58}
\end{align}
 It follows from conditions~\thetag{56} that the numbers~$p_l$
are bounded above and we can choose one of the maximal numbers among them.
 In other words, there exists an index, which will be denoted by~$L$,
such that the following condition is satisfied:
\begin{equation}
p_L\ge p_l \qquad \text{for all}\quad l=0,1,\dotsc.
\tag{59}
\end{equation}

\begin{lemma}
 For an arbitrary sequence of numbers~$p_l$,
$l=0,1,\dots$,
satisfying conditions~\thetag{56}--\thetag{59} the following
inequalities
hold{\rm:}
\begin{gather}
|L-\ol l|\le(3Dl)^{3/4},
\tag{60}
\\
p_L\ge\frac1{\sqrt{27Dl}}.
\tag{61}
\end{gather}
\end{lemma}
***
{\bf Proof.}
 For any
$\Delta>  0$,
the following Chebyshev inequality
holds:
\begin{equation}
\sum_{l=0}^\infty p_l\Theta(|l-\ol l|-\Delta)
\le\sum_{l=0}^\infty p_l\frac{(l-\ol l)^2}{\Delta^2}
=\frac{Dl}{\Delta^2}.
\tag{62}
\end{equation}
 It follows from~\thetag{62}, \thetag{56}, and~\thetag{59} that,
for any
$\Delta>  0$,
we have the chain of inequalities
\begin{equation}
2\Delta p_L
\ge\sum_{|l-\ol l|<\Delta}p_l
\ge1-\frac{Dl}{\Delta^2}.
\tag{63}
\end{equation}
 From~\thetag{63} we obtain
\begin{equation}
p_L\ge\frac1{2\Delta}\biggl(1-\frac{Dl}{\Delta^2}\biggr)
\qquad \text{for any}\quad \Delta>0.
\tag{64}
\end{equation}
 It can be easily shown that the function on the right-hand side
of inequality~\thetag{64} attains its maximum value (with respect
to the variable~$\Delta$)
equal to
$1/\sqrt{27Dl}$
at
$\Delta=\sqrt{3Dn}$.
 Inequality~\thetag{61} is thus proved.
 Now, take
$\Delta=|L-\ol l|$;
then it follows from~\thetag{62} and~\thetag{56} that
\begin{equation}
p_L
\le\sum_{l=0}^\infty p_l\Theta(|l-\ol l|-|L-\ol l|)
\le\frac{Dl}{(L-\ol l)^2}.
\tag{65}
\end{equation}
 Taking~\thetag{61} into account, from~\thetag{65} we obtain
inequality~\thetag{60}.

 To each group there corresponds a particular nonlinear averaging
of the expenditure; as in~\thetag{15}, the averaging of the
expenditure
for FIs from the group with index~$\alpha$
can be expressed by the
formula
\begin{equation}
M_\alpha(\beta,N_\alpha)
=-\frac1{\beta N_\alpha}
\ln\biggl(\frac{N_\alpha!\,(G_\alpha-1)!}{(N_\alpha+G_\alpha-1)!}
\sum_{\{k\}_\alpha}^{N_\alpha}
\prod_{i=i_\alpha}^{j_\alpha}
\bigl(\gamma_i(k_i)e^{-\beta\lambda_ik_i}\bigr)\biggr),
\tag{66}
\end{equation}
where
$\gamma_i(k_i)$
is given by formula~\thetag{12}
and~$\sum_{\{k\}_\alpha}^{N_\alpha}$
denotes the sum over all collections
$\{k\}_\alpha$
of nonnegative
integers
$k_{i_\alpha},\dots,k_{j_\alpha}$
satisfying condition~\thetag{18}.
 As in~\thetag{26} and~\thetag{37}, for all nonnegative
integers~$N_\alpha$,
for all
$\beta>  0$,
and all
$\nu>  -\lambda_1$,
we define the functions
\begin{align}
\Gamma_\alpha(\beta,N_\alpha)
&=\frac{(N_\alpha+G_\alpha-1)!}{N_\alpha!\,(G_\alpha-1)!}
\exp\bigl(-\beta N_\alpha M_\alpha(\beta,N_\alpha)\bigr),
\tag{67}
\\
\zeta_\alpha(\beta,\nu)
&=\sum_{N_\alpha=0}^\infty
\Gamma_\alpha(\beta,N_\alpha)e^{-\beta N_\alpha\nu}.
\tag{68}
\end{align}

\begin{proposition}
 The following relations hold{\rm:}
\begin{align}
\Gamma(\beta,N)
&=\sum_{\{N_\alpha\}}^N\prod_{\alpha=1}^m
\Gamma_\alpha(\beta,N_\alpha),
\tag{69}
\\
\zeta(\beta,\nu)
&=\prod_{\alpha=1}^m\zeta_\alpha(\beta,\nu),
\tag{70}
\\
\zeta_\alpha(\beta,\nu)
&=\prod_{i=i_\alpha}^{j_\alpha}
\frac1{(1-\exp(-\beta(\lambda_i+\nu)))^{g_i}},
\tag{71}
\end{align}
where
$\sum_{\{N_\alpha\}}^N$
denotes the sum over all collections
$\{N_\alpha\}$
of nonnegative integers
$N_1,\dots,N_m$
satisfying
condition~\thetag{19}.
\end{proposition}

{\bf Proof.}
 Relations~\thetag{70} and~\thetag{71} follow from~\thetag{38}.
 To prove~\thetag{69}, it suffices to use formula~\thetag{51} and
take into account the fact that, for all
$\Gamma_\alpha(\beta,N_\alpha)$,
\
$\alpha=1,\dots,m$,
similar formulas hold, and also use the fact
that, in formula~\thetag{51}, the summands of the sum split into
products, which follows from the explicit forms of
$\gamma(\{k\})$~\thetag{13}
and
$\cB(\{k\})$~\thetag{14}.

 The average number of FIs belonging to the group with index~$\alpha$
in the overall purchase of~$N$
 FIs is determined as a function
of~$\beta$
as follows:
\begin{equation}
\ol N_\alpha(\beta,N)
=\frac1{\Gamma(\beta,N)}\sum_{\{k\}}^N
N_\alpha(\{k\})\gamma(\{k\})
\exp\bigl(-\beta\cB(\{k\})\bigr),
\tag{72}
\end{equation}
where
$N_\alpha(\{k\})$
depends on the collection
$\{k\}$
by
formula~\thetag{18}.
 In studying the averages~\thetag{72}, we also
need the following functions:
\begin{align}
\wt N(\beta,\nu)
&=\frac1{\zeta(\beta,\nu)}
\sum_{N=0}^\infty N\Gamma(\beta,N)
e^{-\beta N\nu},
\tag{73}
\\
D\wt N(\beta,\nu)
&=\frac1{\zeta(\beta,\nu)}\sum_{N=0}^\infty
\bigl(N-\ol N(\beta,\nu)\bigr)^2
\Gamma(\beta,N)e^{-\beta N\nu},
\tag{74}
\\
\wt N_\alpha(\beta,\nu)
&=\frac1{\zeta(\beta,\nu)}\sum_{\{k\}}N_\alpha(\{k\})
\gamma(\{k\})\exp\bigl(-\beta(\cB(\{k\})+\nu N(\{k\}))\bigr),
\tag{75}
\\
D\wt N_\alpha(\beta,\nu)
&=\frac1{\zeta(\beta,\nu)}\sum_{\{k\}}
\bigl(N_\alpha(\{k\})-\wt N_\alpha(\beta,\nu)\bigr)^2
\gamma(\{k\})
\exp\bigl(-\beta(\cB(\{k\})+\nu N(\{k\}))\bigr),
\tag{76}
\end{align}
where
$N(\{k\})$
depends on the collection
$\{k\}$
by formula~\thetag{31}
and~$\sum_{\{k\}}$
denotes the sum over all collections
$\{k\}$
of nonnegative integers
$k_1,\dots,k_n$.

\begin{proposition}
 The functions~\thetag{73}--\thetag{76} satisfy the identities
\begin{align}
\wt N(\beta,\nu)
&=\sum_{i=1}^n\frac{g_i}{\exp(\beta(\lambda_i+\nu))-1},
\tag{77}
\\
D\wt N(\beta,\nu)
&=\sum_{i=1}^n\frac{g_i\exp(\beta(\lambda_i+\nu))}
{(\exp(\beta(\lambda_i+\nu))-1)^2},
\tag{78}
\\
\wt N_\alpha(\beta,\nu)
&=\sum_{i=i_\alpha}^{j_\alpha}
\frac{g_i}{\exp(\beta(\lambda_i+\nu))-1},
\tag{79}
\\
D\wt N_\alpha(\beta,\nu)
&=\sum_{i=i_\alpha}^{j_\alpha}
\frac{g_i\exp(\beta(\lambda_i+\nu))}
{(\exp(\beta(\lambda_i+\nu))-1)^2}.
\tag{80}
\end{align}
\end{proposition}

{\bf Proof.}
 First, note that the functions~\thetag{73} and~\thetag{74} can
be written as
\begin{align}
\wt N(\beta,\nu)
&=-\frac1\beta\frac\partial{\partial\nu}
\ln\bigl(\zeta(\beta,\nu)\bigr),
\tag{81}
\\
D\wt N(\beta,\nu)
&=-\frac1\beta\frac\partial{\partial\nu}\wt N(\beta,\nu);
\tag{82}
\end{align}
this can be verified by differentiating
$\zeta(\beta,\nu)$~\thetag{37}.
 Next, substituting~\thetag{38} into~\thetag{81} and~\thetag{82},
we obtain~\thetag{77} and~\thetag{78}.
 Now, taking into account
the explicit form of
$\gamma(\{k\})$~\thetag{13} and
$\cB(\{k\})$~\thetag{14}
as well as formulas~\thetag{66}, \thetag{67}, and~\thetag{70},
we can write expressions~\thetag{75}, \thetag{76} in a form similar
to~\thetag{73}, \thetag{74}:
\begin{align}
\wt N_\alpha(\beta,\nu)
&=\frac1{\zeta_\alpha(\beta,\nu)}
\sum_{N_\alpha=0}^\infty N_\alpha\Gamma_\alpha(\beta,N_\alpha)
e^{-\beta N_\alpha\nu},
\tag{83}
\\
D\wt N(\beta,\nu)
&=\frac1{\zeta_\alpha(\beta,\nu)}\sum_{N_\alpha=0}^\infty
\bigl(N_\alpha-\ol N_\alpha(\beta,\nu)\bigr)^2
\Gamma_\alpha(\beta,N_\alpha)e^{-\beta N_\alpha\nu}.
\tag{84}
\end{align}
 Therefore, the functions~\thetag{75} and~\thetag{76} satisfy formulas
similar to~\thetag{81} and~\thetag{82}, only
$\zeta(\beta,\nu)$
must be replaced by
$\zeta_\alpha(\beta,\nu)$;
substituting~\thetag{71} into these formulas, we obtain~\thetag{79}
and~\thetag{80}.
 The proposition is proved.

 In order to prove the exponential estimate in the assertion of
Theorem~2,
we shall use the following assertion.

\begin{proposition}
 For all positive~$\Delta$,
the following inequality holds{\rm:}
\begin{align}
&
\frac1{\zeta(\beta,\nu)}\sum_{\{k\}}
\Theta\biggl(\sum_{\alpha=1}^m
|N_\alpha-\wt N_\alpha(\beta,\nu)|-\Delta\biggr)\gamma(\{k\})
\exp\bigl(\beta(\cB(\{k\})+\nu N(\{k\}))\bigr)
\tag{85}
\\ &\qquad
\le2^m\exp\biggl(-\frac{\Delta^2}{2Gd}\biggr),
\notag
\end{align}
where
$d$
is defined by formula~\thetag{28}.
\end{proposition}

{\bf Proof.}
 We use the following properties of the hyperbolic cosine
$\cosh(x)=(e^x+e^{-x})/2$:
\begin{equation}
\prod_{\alpha=1}^m\cosh(x_\alpha)
\ge\biggl(\cosh\biggl(\frac\Delta m\biggr)\biggr)^m
\quad \forall x_\alpha:
\sum_{\alpha=1}^m|x_\alpha|\ge\Delta,
\qquad
\frac1{\cosh(x)}\le2e^{-x}.
\tag{86}
\end{equation}
 Inequalities~\thetag{86}, together with formula~\thetag{69},
lead to the following (exponential) Chebyshev inequality:
\begin{align}
&
\sum_{\{k\}}\Theta\biggl(\sum_{\alpha=1}^m
|N_\alpha-\wt{N}_\alpha(\beta,\nu)|-\Delta\biggr)\gamma(\{k\})
\exp\bigl(\beta(\cB(\{k\})+\nu N(\{k\}))\bigr)
\tag{87}
\\ &\qquad
\le2^me^{-c\Delta}\prod_{\alpha=1}^m\biggl(\sum_{N_\alpha=0}^\infty
\Gamma_\alpha(\beta,N_\alpha)e^{\beta\nu N_\alpha}
\cosh\bigl(c(N_\alpha-\wt{N}_\alpha(\beta,\nu))\bigr)\biggr),
\notag
\end{align}
where
$c$
is an arbitrary positive number.
 Taking~\thetag{68}
and~\thetag{70} into account, we rewrite~\thetag{87} in the form
\begin{align}
&
\frac1{\zeta(\beta,\nu)}\sum_{\{k\}}\Theta\biggl(\sum_{\alpha=1}^m
|N_\alpha-\wt{N}_\alpha(\beta,\nu)|-\Delta\biggr)\gamma(\{k\})
\exp\bigl(\beta(\cB(\{k\})+\nu N(\{k\}))\bigr)
\\ &\qquad
\le2^me^{-c\Delta}\prod_{\alpha=1}^m\frac1{2\zeta_\alpha(\beta,\nu)}
\biggl(\exp(-c\wt{N}_\alpha(\beta,\nu))
\zeta_\alpha\biggl(\beta,\nu+\frac c\beta\biggr)
\tag{88}
\\ &\qquad \kern30mm
+\exp(c\wt{N}_\alpha(\beta,\nu))
\zeta_\alpha\biggl(\beta,\nu-\frac c\beta\biggr)\biggr).
\notag
\end{align}

 Further, we use the identity
\begin{align}
&
\ln\biggl(\zeta_\alpha\biggl(\beta,\nu+\frac c\beta\biggr)\biggr)
=\ln(\zeta_\alpha(\beta,\nu))
+\frac c\beta\frac\partial{\partial\nu}
\ln(\zeta_\alpha(\beta,\nu))
\tag{89}
\\ &\qquad
+\int_\nu^{\nu+c/\beta}d\nu'\,\biggl(\nu+\frac c\beta-\nu'\biggr)
\frac{\partial^2}{{\partial\nu'}^2}\ln(\zeta_\alpha(\beta,\nu')),
\notag
\end{align}
which holds because the function~\thetag{71} can be differentiated
twice with respect to the variable~$\nu$.
 From the explicit
form of the function
$\zeta_\alpha(\beta,\nu)$~\thetag{71}, we
obtain
\begin{equation}
\frac{\partial^2}{\partial\nu^2}\ln(\zeta_\alpha(\beta,\nu))
=\beta^2\sum_{i=i_\alpha}^{j_\alpha}
\frac{g_i\exp(-\beta(\lambda_i+\nu))}
{(\exp(-\beta(\lambda_i+\nu))-1)^2}
\le\beta^2G_\alpha d,
\tag{90}
\end{equation}
where
$d$
is expressed by~\thetag{28}.
 Using inequality~\thetag{90}
and formulas~\thetag{79} and~\thetag{81}, from~\thetag{89} we
obtain the inequality
\begin{equation}
-c\wt N_\alpha(\beta,\nu)
+\ln\biggl(\zeta_\alpha\biggl(\beta,\nu+\frac c\beta\biggr)\biggr)
-\ln(\zeta_\alpha(\beta,\nu))
\le\frac{G_\alpha}2dc^2.
\tag{91}
\end{equation}
 By~\thetag{91} and~\thetag{20}, from~\thetag{88} we obtain the
inequality
\begin{align}
&
\frac1{\zeta(\beta,\nu)}\sum_{\{k\}}\Theta\biggl(\sum_{\alpha=1}^m
|N_\alpha-\wt N_\alpha(\beta,\nu)|-\Delta\biggr)\gamma(\{k\})
\exp\bigl(\beta(\cB(\{k\})+\nu N(\{k\}))\bigr)
\tag{92}
\\ &\qquad
\le2^m\exp\biggl(-c\Delta+\frac{Gdc^2}2\biggr).
\notag
\end{align}
 The expression on the right-hand side of inequality~\thetag{92}
contains an arbitrary positive parameter~$c$;
it attains its
minimum equal to
$2^m\exp(-\Delta^2/2d)$
at
$c=\Delta/d$,
and
hence inequality
\thetag{85}
is a consequence of~\thetag{92}.
 The proposition is proved.

 Let us suppose that~$\lambda_1$
and~$\lambda_n$
are independent of
$N$;
by~\thetag{10}, this implies that the prices~$\lambda_i$,
\
$i=1,\dots,n$,
lie in an interval independent of
$N$.
 The form of the dependence of~$n$
on~$N$
is not specified, since it does not affect the results
obtained.
 It should only be noted that, by~\thetag{11}, we have
$n\le  G$.
 The conditions formulated above essentially
restrict neither the choice of~$\lambda_i$
and~$g_i$ nor that of the partition of FIs into groups.
 Because of this, we cannot
write the asymptotics of the nonlinear averages~\thetag{15}
and~\thetag{66}
and of the averages~\thetag{72} in the limit as
$N\to  \infty$
in the general case; however, it turns out that we can prove
some properties of these averages in such a limiting case.

{\bf Proof of Theorem~2.}
 By Proposition~4, for given~$\beta$
and~$N$,
we choose
a~$\nu'$
such that inequality~\thetag{53} holds.
 By formulas~\thetag{37}, \thetag{38}
and~\thetag{73}, \thetag{74}, \thetag{77}, \thetag{78}, the sequence
$$
p_l=\frac{\Gamma(\beta,l)e^{\beta l\nu'}}{\zeta(\beta,\nu')}
$$
is a sequence for which Lemma~2 is applicable.
 This means that
for given~$\beta$,
$N$,
and the corresponding~$\nu'$,
the following inequalities hold:
\begin{gather}
|\wt N(\beta,\nu')-N|
\le\bigl(3D\wt N(\beta,\nu')\bigr)^{3/4},
\tag{93}
\\
1>\frac{\Gamma(\beta,N)e^{\beta N\nu'}}{\zeta(\beta,\nu')}
\ge\frac1{\sqrt{27D\wt N(\beta,\nu')}}.
\tag{94}
\end{gather}

 Now, note that the function
$\wt N(\beta,\nu)$~\thetag{77} possesses
the following properties for all
$\nu>  -\lambda_1$
if
$\beta<  0$
and all
$\nu<  -\lambda_n$
if
$\beta>  0$:
\begin{gather}
\wt N(\beta,\nu)\ge0,
\qquad
\lim_{\nu\to\infty}\wt N(\beta,\nu)=0,
\qquad
\lim_{\nu\to-\lambda_1+0}\wt N(\beta,\nu)
=\lim_{\nu\to-\lambda_n-0}\wt N(\beta,\nu)
=+\infty,
\tag{95}
\\
\frac1\beta\frac{\partial\wt N}{\partial\nu}(\beta,\nu)
=\sum_{i=1}^n\frac{g_i\exp(-\beta(\lambda_i+\nu))}
{(\exp(-\beta(\lambda_i+\nu))-1)^2}>0,
\notag
\\
\biggl|\frac{\partial\wt N}{\partial\nu}(\beta,\nu)\biggr|
\le\begin{cases}
-\dfrac{\beta G\exp(-\beta(\lambda_1+\nu))}
{(\exp(-\beta(\lambda_1+\nu))-1)^2}
& \text{for}\ \beta<0, \\
\dfrac{\beta G\exp(-\beta(\lambda_n+\nu))}
{(\exp(-\beta(\lambda_n+\nu))-1)^2}
& \text{for}\ \beta>0.
\end{cases}
\notag
\end{gather}
 By~\thetag{95}, first, equation~\thetag{25} has a solution
$\nu>  -\lambda_1$
for all
$\beta<  0$
and
$\nu<  -\lambda_n$
for all
$\beta>  0$;
we denote this solution by
$\nu(\beta,N)$.
 Second, it follows
from~\thetag{93} and~\thetag{95} that
\begin{equation}
\nu'=\nu(\beta,N)+O\biggl(\frac1{N^{1/4}}\biggr),
\tag{96}
\end{equation}
because, by~\thetag{10},
$D\wt N(\beta,\nu)$~\thetag{78} satisfies
the estimate
\begin{align}
\frac{-\beta G\exp(-\beta(\lambda_n+\nu))}
{(\exp(-\beta(\lambda_n+\nu))-1)^2}
&\le D\wt N(\beta,\nu)
\le\frac{-\beta G\exp(-\beta(\lambda_1+\nu))}
{(\exp(-\beta(\lambda_1+\nu))-1)^2}
\qquad \text{for}\quad \beta<0,
\tag{97}
\\
\frac{\beta G\exp(-\beta(\lambda_1+\nu))}
{(\exp(-\beta(\lambda_1+\nu))-1)^2}
&\le D\wt N(\beta,\nu)
\le\frac{\beta G\exp(-\beta(\lambda_n+\nu))}
{(\exp(-\beta(\lambda_n+\nu))-1)^2}
\qquad \text{for}\quad \beta>0
\notag
\end{align}
and~$G$
increases in the same way as~$N$
by condition~\thetag{21}.
 Let us now take into account the fact that the functions
$\wt N_\alpha(\beta,\nu)$
satisfy identity~\thetag{79}, and, therefore, they possess properties
similar to~\thetag{95}.
 Then it follows from~\thetag{96}, \thetag{79}, \thetag{95}
and the definitions of
$\cN_\alpha(\beta,N)$~\thetag{24} that
\begin{equation}
|\cN_\alpha(\beta,N)-\wt N_\alpha(\beta,\nu')|
=O(N^{3/4}).
\tag{98}
\end{equation}

 Next, suppose that
$\Delta=aN^{3/4+\delta}$,
where
$a$
and~$\delta$
are arbitrary positive parameters independent of
$N$.
 If the condition
$$
|\wt N_\alpha(\beta,\nu')-N_\alpha|\ge\Delta,
$$
is satisfied, then the following condition is satisfied for
sufficiently large~$N$:
$$
\sum_{\alpha=1}^m|\cN_\alpha(\beta,N)-N_\alpha|
\ge\Delta'
=\Delta-\sum_{\alpha=1}^m
|\cN_\alpha(\beta,N)-\wt N_\alpha(\beta,\nu')|.
$$
 This yields the inequality
\begin{align}
&
\sum_{\{k\}}\Theta\biggl(\sum_{\alpha=1}^m
|N_\alpha-\cN_\alpha(\beta,\nu)|-\Delta\biggr)\gamma(\{k\})
\exp\bigl(\beta(\cB(\{k\})+\nu N(\{k\}))\bigr)
\tag{99}
\\ &\qquad
\le\sum_{\{k\}}\Theta\biggl(\sum_{\alpha=1}^m
|N_\alpha-\wt N_\alpha(\beta,\nu')|-\Delta'\biggr)\gamma(\{k\})
\exp\bigl(\beta(\cB(\{k\})+\nu N(\{k\}))\bigr).
\notag
\end{align}
 In view of~\thetag{98} and~\thetag{43}, the following estimate
holds for any parameter
$\varepsilon>  0$
independent of
$N$
and for all positive~$a$
and~$\delta$
as
$N\to  \infty$:
\begin{equation}
\sqrt N\exp\biggl(-\frac{{\Delta'}^2}{2Gd}\biggr)
=O\biggl(\exp\biggl(-\frac{(1-\varepsilon)a^2N^{1/2+2\delta}}
{2\wt gd}\biggr)\biggr).
\tag{100}
\end{equation}
 Relations~\thetag{85}, \thetag{94}, and~\thetag{100}
yield~\thetag{27}.
 The theorem is proved.

 Here we have studied the case
$\beta>  0$.
 In exactly the
same way, we may consider the case
$\beta<  0$.
 If we take
$\nu<  -\alpha_n$ everywhere, then we shall obtain the
same results.

\section{The tunnel canonical operator in economics}
 Equilibrium prices are determined from the condition of
the equality between supply and demand in each commodity and
resource.
 Similarly, the following pairs are determined: ``flows of commodities
and services--prices''; ``flows of
labor resources--level of wages''; ``flows of natural
resources--rents''; and ``loan interest rate--loan volume.''

 The asymptotics of
$M$ and~$\wt M$
is given by the tunnel
canonical operator in the phase space of pairs~\cite{5}.

 We consider the phase space~$\mathbb R^{2n}$,
where the intensive
quantities play the role of coordinates and the extensive ones,
of momenta.
 In economics, the role of the values of the random
variable~$\lambda_i$
can be played by the prices of the corresponding commodities
and~$N_i$
is, for example, the number of commodities sold,
i.e., the number of people who have bought of this commodity or
the bank rate of the $i$th
bank, etc.
 Obviously, the price depends on
the demand, i.e.,
$\lambda_i(N_i)$
is a curve in two-dimensional
phase space.
 In two-dimensional phase space, to each point
(vector)~$\lambda_i$,
\
$i=1,\dots,n$,
there corresponds a vector
$N_i(\lambda_1,\dots,\lambda_n)$,
\
$i=1,\dots,n$.
 In a more general case, we deal with an
$N$@-dimensional manifold
(surface), where the ``coordinates'' and the ``momenta'' locally
depend on~$n$
parameters and, moreover, a certain condition
holds: the Lagrange brackets of the ``coordinates'' and of ``the
momenta'' with respect to these parameters are zero.
 Therefore,
the author has called such a manifold {\it Lagrangian}.
 In other
words, we can say that the form
$\sum N_i\,d\lambda_i$
is closed
(see the concluding remarks in~\cite{6} and~\cite{7}).
 Hence
$\int N_id\lambda_i$
is independent of the path and, just as
in mechanics,
$\int p\,dq$
($p$
is the momentum,
$q$
is the coordinate),
can be called an {\it action}.

 The producer buys resources and transforms a resource expenditure
vector into a commodity production vector.
 The consumer buys these
commodities.
 Accordingly, the equilibrium prices for resources
and for consumer goods are determined by the equalities given
above (see also~\cite{8}--\cite{17}).

 In addition to such equilibrium prices, vertical pairs of isolated
consumer goods and pairs ``seller--buyer,'' i.e., ``permanent
seller--permanent buyer,'' can also be formed and the corresponding
prices related to this pairs are generated.
 The analog in quantum
statistics are Cooper pairs.

 This construction requires the use of the ultrasecond quantization
method in an abstract algebraic form which could be applicable
in economics.
 Such a theory leads to the formation of ``vertical''
clusters.
 The method in question is developed in another paper.

\end{document}